\documentstyle[12pt]{article}
\newcommand{\abstitle}[1]{{\small {\bf #1}}}
\newcommand{\absauthor}[1]{\small {\bf #1}}
\newcommand{\address}[1]{{\it #1}}
\newcommand{\ba}{\begin{array}{1}}
\newcommand{\be}{\begin{equation}}
\newcommand{\ee}{\end{equation}}
\newcommand{\ea}{\end{array}}
\newcommand{\bb}{\begin{thebibliography}}
\newcommand{\eb}{\end{thebibliography}}
\newcommand{\ci}[1]{\cite{#1}}
\newcommand{\bi}[1]{\bibitem{#1}}

\newcommand{\ra}{\rightarrow}

\begin{document}
\begin{center}
\abstitle{SEMIHARD HADRON PROCESSES \\
AND QUARK-GLUON STRING MODEL}
\footnote{encouraged and supported by Russian Foundation of Fundamental
          Research } \\[3.0mm]
\absauthor{ G.I.Lykasov$^a$ and M.N. Sergeenko$^b$, }
\\[2.0mm]
\address{$^a$ Joint Institute for Nuclear Research, Dubna 141980,
Russia}\\ \address{$^b$ Institute of Physics of the Belarus Academy
of Science, Minsk 220602, Belarus}
\end{center}
\begin{abstract}

A new approach to the analysis of soft and semihard hadron processes is
suggested. In the frame of the Quark-Gluon String Model the
interaction of valence quarks and diquarks and sea quarks
(antiquarks) of colliding hadrons is taken into account. This one is
calculated as the exchange by one nonpertubative gluon, i.e., the
cut-off parameter in the gluonic propagator is included. This one
allows us to analyze the inclusive hadron spectra in hadron
collisions at transverse momenta up to $3-4$ Gev/c.   \end{abstract}

\section{Introduction}
%{\bf 1 Introduction}
As it is known the quark-gluon strings model (QGSM) [1-6] based on the $1/N$
expansion in QCD [7-9] had a significant success by the description of
different hadron characteristics. For example, it reproduces a large amount
of existing experimental data about hadron production at high energies
[1-6].

The QGSM has been successfully applied to the production of
hadrons containing light $u$, $d$ and $s$ quarks. This model
reproduces in great detail and in terms of very few parameters a
large amount of existing  experimental data about hadron production
at high energies in hadron-hadron and  hadron-nucleus collisions
[1-3]. Besides QGSM describes adequately the production of charmed
mesons at the explored low and moderate energies.

However the characteristics
integrated over transverse momentum $p_t$ or at average
transverse momentum $<p_t>$ are considered only in the
framework of this model. So this model is limited usually by the
analysis of soft hadron reactions or processes at small transverse
momenta $p_t$ of produced hadrons.  There are some versions of QGSM
taking into account transverse momenta of quarks in the  initial
hadrons \cite{10,11,Lyk} and \ci{Ranft}.  It allowed to describe
inclusive spectra of hadrons produced in $hN$ collisions up to $p_t =
1.-1.5$ Gev/c.  But there was a large sensitivity of these
ones to the initial quark distribution over transverse momentum in a
hadron which has been parametrised usually in an exponential form. On
the other hand at large $p_{t}$ the ordinary perturbative QCD can be
applied to the analysis of hard processes.  Semihard ones can be
explained by QCD models of type as \cite{Lev,Nas} etc. However the
question arises whether one can apply the QGSM to the analysis of
semihard processes if to take into account the colour interaction
between colliding quarks (antiquarks) and diquarks (quarks) before the
creation and dacay of $q\bar q$ (or $q(qq)$) string.

So the  phenomenological Pomeron exchange corresponding to the
cylinder graph can be understood in the framework of QGSM [1-3] as
the exchange of two gluons \cite{12,13}.  However the perturbative
calculation of the elastic quark-quark scattering amplitude through
two-gluon exchange shows a singularity at $t=0$. Although this
singularity can be cancelled when incorporating quarks in the proton
wave function, this procedure is not able to reproduce the
$t$-dependence of the differential cross section observed in
experiment \ci{14}. Meanwhile some time ago the Pomeron was
described by the exchange of two nonpertubative gluons \cite{15,16,17}, where
by nonpertubative it means a gluon whose propagator does not show a
pole at $k^2=0$. In \ci{18} the two-gluon exchange model was applied
to the description of nucleon-nucleon scattering where a gluon
propagator regularized by a dynamically generated gluon mass, derived
by Cornwall some years ago \ci{19}.

The production of hadrons at high energies is described in the QGSM
by "cutting" the forward scattering diagrams of the "cylindrical"
type [1-3]. Each cylinder corresponds to the exchange of a single
Pomeron.  Such Pomeron exchange as the one of two non-perturbative
gluons can be considered for the calculation of cylinder-cut
graphs in the framework of the QGSM. This paper is dedicated to the
exploitation of this idea  to the analysis of hadron processes in the
framework of QGSM.

\section{General formalism} %{\bf 2 General formalism}

Consider  the inclusive spectrum of hadrons produced in
nucleon-nucleon collision in the framework of QGSM. As is known the
main contribution to such processes at large energies gives the
graphs of cylinder-cut type in the $s$ channel  corresponding to the
Pomeron exchange in $t$ channel [1-6].
The hadron production can be considered in the
following manner \cite{11}. Each of two colliding nucleons is
divided into a quark and a diquark with the opposite transverse
momenta.  After the colour interaction (by means of
non-perturbative gluon) between the quark of first proton and
diquark of another one, and diquark of the first proton and
quark of second one  two quark-gluon strings  are created in the
chromostatic constant field; then these two strings decay into
secondary hadrons.  This process is repeated $n$ times during the
production of $n$ Pomeron showers (or 2n $q\bar q$ chains). Such
division of the transverse momentum considered in \ci{11} is
analogous to the so called consequent division of energy between
$n$ Pomeron showers in the multiperipheral model of the hadron
production \ci{2}.

In principle, there can be so called regular  division of energy and
transverse momentum between 2n $q\bar q$ chains or $n$ Pomeron
showers. But at all versions of QGSM nobody considered the colour
interaction between quark an antiquark or diquark and quark
respectively. For the soft processes, i.e., at small $p_t$, quark
distributions over $x$ on the ends of $q-\bar q$ strings are found
from Regge asymptotic of graphs (cylinder or planar type) at $x\ra 1$
[1-5].  The decay of each this string into hadrons is described by
fragmentation functions, the form of which is determined by the Regge
asymptotic of these graphs too [1-5]. The dependence of hadron
inclusive spectra on $p_t$ is the result of the inclusion of
internal hadron transverse momenta of quarks only \cite{10,11} and
\cite{Ranft}.  This approach is good to be applied to hadron
production at $p_t\leq 1.-1.5$ GeV/c \cite{10,11,Lyk}.

However if we want to analise the hadron production processes at not
large $p_t$, for example, $p_t\leq 3-4$ Gev/c, the question
arises whether one can use the QGSM for this aim. In order to do it
we must know the distribution of quarks (antiquarks) and diquarks
over transverse momenta $k_t$ on the ends of $q\bar q$ string after
some colour interaction between constituents of colliding hadrons.
For this aim one can use the approach suggested in \ci{18} where the
Pomeron is described by the exchange of two nonperturbative gluons and
it means that a gluon propagator has not a pole at $k^2=0$ \ci{16}.
For example, in \ci{17} an approximate Shwinger-Dayson [SD] equation for
the gluon propagator in the axial gauge has been solved and found a
solution whose infrared behaviour is less singular than a pole at
$k^2=0$.  This approach has given reasonable agreement with
experimental data. Recently in \ci{18} it has been shown how a gluon
propagator regularized by a dynamically generated gluon mass, derived
in \ci{19} some years ago, successfully describes nucleon-nucleon
scattering in two-gluon exchange model. The gluon propagator
represented as an approximate solution of the SD equation and it had
not the infrared singularity at $k^2=0$ as a result of a some cut-off
parameter $m_0$ or so called "effective" gluon mass. In fact the
model of ref. \ci{18} has a single  parameter which was taken to be
the ratio of $m_0$ and $\Lambda=\Lambda_{QCD}$.

In this paper we take into account the interaction between quark and
antiquark (or diquark and quark) before the creation and the decay of the
corresponding string. The main contribution to this one results
in the one gluon-exchange between them, which can be calculated using
the gluon propagator with the cut-off parameter as like as in \ci{18,19}.
So this interaction can be considered as the
elastic $q\bar q$ or quark-diquark ($q(qq)$) scattering calculated in
the one-gluon exchange approximation when transverse momenta of
quarks or diquarks are changed only. If one represents the
distribution of quarks or diquarks in a nucleon over $x$ and $p_t$ in
the factorized form, i.e.
\be
f_{\tau}(x,\vec{k}_t)=f_{\tau}(x)g_{\tau}^{(0)}(\vec{k}_t),
\ee
where $\tau$ means the flavour of a quark, $x$ and $\vec{k}_t$ are
its fraction of the longitudinal momentum and the transverse one
respectively; $f_{\tau}(x)$ is the $x$-distribution and
$g_{\tau}^{(0)}(\vec{k}_t)$ is the $k_t$-distribution of this quark
(antiquark or diquark) in the initial hadron. Then the factorized
form of $f_{\tau}(x,\vec{k}_t)$ will be true after the one-gluon
exchange, but the distribution $g_{\tau}^{(0)}(\vec{k}_t)$ will be
changed only. After such interaction this one can be
calculated by the following manner:
\be
g_{\tau}^{(1)}(\vec{k}_{1,t}) = C_1\int D^2(q^2)
g_{\tau}^{(0)}(\vec{k}_t) \delta^{(2)}(\vec{k}_{1,t} - \vec{k}_t -
\vec{q}_t)d^2q_t\,d^2k_t
\ee
or
$$ g_{\tau}^{(1)}(\vec{k}_{1,t}) = C_1\int D^2[(\vec{k}_{1,t} -
\vec{k}_{t})^2]g_{\tau}^{(0)}(\vec {k}_t) d^2k_t $$
where $D(q^2)$ is the regularized gluon propagator and according to
\ci{18,19} it can be written in the following form:
\be
D(q^2)=\frac{\alpha_s(q^2)}{q^2+m^2(q^2)}
\ee
where:
\be
\alpha_s(q^2)=\frac{1}{b_0ln\frac{q^2+4m^2(q^2)}{\Lambda^2}},
\ee
\be
m^2(q^2) =
m_0^2\left[\frac{ln((q^2+4m_0^2)/\Lambda^2)}{ln(4m_0^2/\Lambda^2)}
\right]^{-12/11};
\ee
Here in (4), (5) $b_0=(33-2n_f)/48\pi^2$; $m_0=0.37$ Gev for
$\Lambda=0.3$ Gev. As in our previous paper \ci{11,Lyk} we choose
the  quark $p_{t}$-distribution in the initial hadron  in
the Gauss form normalized to $1$,
\be
g_{0}(\vec{k}_{t}) = \frac{\gamma}{\pi}exp(-\gamma \vec
{k}^{2}_{t}).
\ee
with $\gamma $ to be a free parameter, $\gamma=1/<k^2_t>$ where $<k^2_t>$ is
the average value of the squered transverse momentum of quarks in a nucleon.
The  constant $C_1$ in (2) is
determined by the normalization condition:
\be
\int{g_{\tau}^{(1)}(\vec{k}_t)d^2k_t}=1.
\ee
As it is seen from (3) the gluon propagator, $D(q^2)$,  at very large
$q^2$ returns to the usual one asymptotically, i.e.
$D(q^2)\propto q^{-2}$, because $m^2(q^2) \ra 0$ at $q^2\ra \infty $.
This method of the calculation of the $k_t$-distribution of a
quark (antiquark or a diquark) on the ends of $q\bar q$ string allows
us to construct inclusive spectra of hadrons produced from its decay
as the function of $x$ and $p_t$. Note the sensitivity of $g_{\tau}^{(1)}$
calculated using (2) to the choice of the $g_{\tau}^{(0)}$ form is very
small if we consider not small $k_t$. It is caused by the strong dependence
of the gluon propagator (3) on $q_t$.

Let us return to this construction in the framework of  the version
of QGSM taking into account transverse momenta of quarks \ci{11,Lyk}. The
expression for the invariant inclusive hadron spectrum for the
reaction $pp\ra hX$  corresponding to graphs of the
cylinder-cut type [1-6] can be written in the following form \ci{11}:
\be
\rho _{pp\ra hX}(x,\vec{p}_t)\equiv
E\frac{d\sigma_A}{d^3\vec{p}} = \sum_{n=1}^{\infty}\sigma_n(s)
\phi_n(x,\vec{p}_t),
\ee
where $\sigma_n$ is the cross section for  production of the
$n$ pomeron chain (or $2n$ quark-antiquark strings) decaying into
hadrons, calculated in the frame of the "eiconal approximation"
\cite{20}, $\phi_n(x,\vec{p}_t)$ is the $x$ and $p_t$-distributions
of hadrons produced in the decay of the $n$ pomeron chain. These
functions, $\phi_n(x,p_t)$, were represented in the following form
\ci{11}:
\be
\phi_n(x,\vec{p}_t)=\int_{x_+(n)}^1dx_1\int_{x_-(n)}^1dx_2
\Psi_n(x_n,\vec{p}_t;x_1,x_2) ;
\ee     with
\begin{eqnarray}
\Psi_n(x_n,p_t;x_1,x_2)= \nonumber
F_{qq}^{(n)}(x_+(n),\vec{p}_t;x_1)F_{q_v}^{(n)}(x_-(n),\vec{p}_t;x_2)/\bar
{F_{q_v}}^{(n)}(0,\vec{p}_t)+ \\ \nonumber
F_{q_v}^{(n)}(x_+(n),\vec{p}_t;x_1)F_{qq}^{(n)}(x_-(n),\vec{p}_t;x_2)/\bar
{F_{qq}}^{(n)}(0,\vec{p}_t)+ \\ \nonumber
2(n-1)F_{q_{sea}}^{(n})(x_+(n),\vec{p}_t;x_1)F_{\bar
{q_{sea}}}^{(n)}(x_-(n),\vec{p}_t;x_2) ; \nonumber
\end{eqnarray}    where
$x_{\pm}=x_{\pm}(n)=0.5(\sqrt{x_t^2+x^2_n} \pm x)$,  $x_t=2\sqrt{(m_h^2 +
\vec{p}_t^2)/s}$ ; $m_h$ is the mass of the produced hadron, $s$ is
the total energy squared of colliding protons in c.m.s. Note
that Feynman variable $x$ in $n$-Pomeron shower depend on $n$
\ci{11}: $x_n = x_/(1-x_0)^{n-1}$,\,\, $x_0 \simeq 0.25$.

The functions $F_{\tau }^{(n)}(x_{\pm}(n),\vec{p}_t)$ are represented
by convolutions
\be
F_{\tau}^{(n)}(x_{\pm}(n),\vec{p}_t;x_{1,2})=
\nonumber \int d^2k_t f_{\tau}^{(n)}(x_{1,2},\vec{k}_t)G_{\tau\ra
h}(x_{\pm}(n)/x_{1,2},\vec{k}_t;\vec{p}_t),
\ee
\be
F_{\tau}^{(n)}(0,\vec{p}_t;x_{1,2})= \nonumber \int
d^2k_t f_{\tau}^{(n)}(x_{1,2},\vec{k}_t)G_{\tau\ra h}(0,\vec{p}_t)
\nonumber = G_{\tau\ra h}(0,\vec{p}_t).
\ee
Here $\tau$ means the flavour of the valence (or sea quark) or
diquark, $f_{\tau}^{(n)}(x,\vec{k}_t)$ is the quark distribution
function, depending on the longitudinal momentum fraction $x$ and the
transverse momentum $\vec{k}_t$ in the $n$-Pomeron chain:
\be
G_{\tau\ra h}(z,\vec{k}_t;\vec{p}_t)=zD_{\tau\ra
h}(z,\vec{k}_t;\vec{p}_t), \nonumber
\ee
$D_{\tau\ra h}(z,\vec{k}_t;\vec{p}_t)$ is the fragmentation function
of a quark or diquark of flavour $\tau$ into a hadron $h$.

The functions  $f_{\tau}^{(n)}(x)$ and $g_{\tau}^{(n)}(\vec{k}_t)$
can be calculated by using regular \ci{10} or consequential \ci{11}
division of $x$ and $k_t$ between 2n $q\bar q$ chains. Here we use
the second one \ci{11}, so the $g_{\tau}^{(n)}(\vec{k}_t)$ are
calculated according to \ci{11} by the following manner.  The quark
functions $f_{\tau}(x,\vec {k}_t)$ in initial hadrons are represented
in the factorized form (1).
The distributions functions over transverse momentum $p_{t}$,
$g_{\tau}^{(n)}(\vec{k}_{n,t})$, are calculated analogous to (2).
This means that the quark distribution after one nonperturbative gluon
exchange is given by (2).  After exchange by the second gluon the
quark distribution in corresponding chain will be expressed
already via the function $g_{\tau}^{(1)}(\vec{k}_t)$, i.e.:
\be
g_{\tau}^{(2)}(\vec{k}_{2,t}) = C_2\int g_{\tau}^{(1)}(\vec{k}_{1,t})
 D^2[(\vec{k}_{2,t} - \vec{k}_{1,t})^{2}]d^{2}k_{1,t }.
\ee
Repeating this iteration  procedure we obtain the quark distribution
function in the $n$ chain expressed via the function
$g_{\tau}^{(n-1)}(\vec{k}_{n-1,t})$ and therefore via the function
$g_{\tau}^0(\vec{k}_t)$:
\begin{eqnarray}
g_{\tau}^{(n)}(\vec{k}_{n,t}) =  C_n
\int g_{\tau}^{(n-1)}(\vec{k}_{n-1,t}) D^2[(\vec{k}_{n,t }-\vec{k}_{n-1,t
})^{2}] d^{2}k_{n-1,t} =  \\ \nonumber \int d^{2}k_{n-1,t}
D^{2}[(\vec{k}_{n,t} - \vec{k}_{n-1,t })^{2}]\int d^{2}k_{n-2,t }
D^{2}[(\vec{k}_{n-1,t } - \vec{k}_{n-2,t })^{2}]...\\ \nonumber
\cdot \int d^{2}k_t D^{2}[(\vec{k}_{1,t } -
\vec{k}_t)^{2}]g_{\tau}^(0)(\vec{k}_t).
\end{eqnarray}

It is easily to see that in the $n$-pomeron chain the quark
functions will be factorized too:
\be
\tilde{f}^{(n)}_{\tau}(x_n,\vec{k}_{n,t }) =
f^{(n)}_{\tau}(x_n)g^{(n)}_{\tau}(\vec{k}_{n,t }),
\ee
This calculation of $k_t$-distribution of quarks on the ends of $2n$-th
sring is a principal difference from our previous papers \ci{11,Lyk}.
But $x$-distribution after $n$-Pomeron exchanges $f_{\tau}^n(x)$ is
calculated according to \ci{11}, i.e. the Feynman variable $x$ in
$n$-Pomeron shower depend on $n$ (see above).
%\be %g_{\tau}^{(n)}(\vec{k}_t) =
%\int{g_{\tau}(k_{1,t})g_{\tau}(k_{2,t})...g_{\tau}(k_{n,t})
%\delta(k_t-\sum_{i=1}^{i=n}k_{i,t})\prod_{i=1}^{i=n}}d^2k_{i,t}
%\ee
%where each function $g_{\tau}(k_{i,t})$ is calculated using
%expression (2).

The functions $\tilde{G}_{\tau \ra h}(z,\vec{k}_t;
\vec{p}_{t})$ have  represented  in  the  following form \ci{11}:
\be
\tilde{G}_{\tau \ra h}(z,\vec{k}_t; \vec{p}_t) =
G_{\tau \ra h}(z,\vec{p}_t)\tilde{g}_{\tau \ra h}
(\tilde{k}_t),
\ee where
\be
\tilde {g}_{\tau \ra h}(\tilde
{k}_t) = \frac{\tilde{\gamma}}{\pi}exp(-\tilde{\gamma}\tilde
{k}^2_t),
\ee
\be \tilde {k}_t = \vec {p}_t - z\vec{k}_{t },  \ \ \ \ \ \ \
z = {x_{\pm }\over x_{1,2}}.
\ee
Substituting  now  the  functions (14)-(17) to (10) and
integrate it over $d^{2}k_{t }$.
In principle, it can be made by analitically if the function
$g_{\tau}^{(n)}(\vec{k}_t)$ calculated by (10) is approximated by a sum of
exponantials over $k^2_t$ (see Appendix).

%{\bf 3 Results and discussions}
\section{Results and discussions}

The results of calculation of  differential  cross  sections
performed with the help of the modified QGSM are  presented  in
Figs.  1-5.  Using (8) we can calculate different characteristics of
hadrons produced in N-N collisions, for example, correlation of the
average transverse momentum $<p_t>$ and hadron multiplicity $N$
(see for example \cite{11}).
Besides the inclusive spectra of hadrons defined by (8), we can
calculate the distributions of hadrons over $x$ or $p_t$,i.e.:
\be
F(x)=\int{\rho_{pp\ra hX}(x,p_t)d^2p_t}
\ee and
\be
\frac{d\sigma}{dp_t^2}=\frac{\pi}{2}\sqrt{s} \int{\varrho_{pp\ra
hX}(x,p_t)\frac{dx}{E^{*}}}
\ee
where $E^*$ is the energy of final hadrons in N-N c.m.s.

The results of the calculations of $<p_t>(N)$ for $pp$ collisions
are shown in Fig. 1. The calculations reproduce the data at small
multiplicities $N$ and go below at large $N$. The discrepancy
between the results of the calculations and the experimental
data about  $<p_t>(N)$ may be connected with the fact that the
expressions for the cross sections $\sigma_n$ (see (8)) were obtained
with inclusion of only non-enhanced graphs of Reggeon theory [2].
Inclusion of enhanced-type diagrams leads to the appearance of $1/x$
terms in the distributions for sea quarks, whose contribution is
especially large for $x\simeq 0$.

The calculation results for inclusive hadron spectra  performed using
the formulae (8), (20) are presented in Figs. 2-5.  In Figs.
2,\,3 we show the  inclusive spectra of  $\pi ^+$ and $\pi ^-$ mesons
produced in $pp$ collisions at $p_{lab}=200$ GeV/c as a function of
$p_t$ for three values of the Feynman variable x: $x=0$,\, $0.3$\,
and $0.6$. We see a good description of experimental data with
modified QGSM. Corresponding parameter values in formulas (6) and (17)
are: $\gamma = 9$\,(GeV/c)$^{-2}$
what corresponds to $<k^2_t>=1/\gamma\simeq 0.1$\,(GeV/c)$^{2}$,
$\tilde {\gamma}= 7$\,GeV$^{-2}$.
Note that the calculation results depend
on the value of this parameters very weakly.

In Fig. 4 we show the invariant cross section $E(d\sigma /d^3\vec
{p}$ versus transverse momentum $p_t$ for the  $\pi ^0$ for mesons
produced in $pp$ collisions at  $\sqrt{s}=52.7$ GeV. From this
figure it can be seen that our version of the QGSM give good
description of the data [27] up to $p_t\,<\,3$ GeV/c. Some discrepancy
with data at larger $p_t$ can be caused by the following.
First of all
the $p_t$ quark distribution in initial hadrons has been taken
in the simple Gauss form (6) which is true at small $p_t$. Second
reason can be conjugated with the Regge trajectories, $\alpha (t)$,
which are taken in the QGSM at $t = 0$. We suppose that in the
calculations it is needed to take into account the dependence $\alpha
(t)$ on transfer momentum $t$.

In Fig. 5 we present the  calculation results for the  differential
cross sections $d\sigma /dp_t^2$ of the reactions $pp\ra D^nX$,
\,$n\,=\,+\,(a),\,-\,(b),\,0\,(c),\,\bar 0\,(d)$, at  $\sqrt{s}=27.4$
GeV.  The curves correspond to the case  when the Regge $\Psi $-
trajectory has intercept, $\alpha _{\Psi }(O)=0,$, which corresponds
to the nonlinear $\Psi $-trajectory.  The respective values of
parameters in fragmentation functions are \cite{Lyk}:
$a_{0}=0.1\,10^{-3}$, $a_{1}=5$.  From this  figure it can be seen
that the version of the QGSM under consideration give good
description of the data [28].

The inclusive spectra of all $D$-mesons have been calculated
with the parameter value $\tilde{\gamma }$, $\tilde{\gamma }=2$
(Gev/c)$^{-2}$.  The  better  coincidence  of  our calculations
with the experimental data is for $D^+$  and $D^0$ mesons.

\section{Conclusion}

In this work we have discussed one possible modification of the
Quark-Gluon String Model to the description of semihard processes.
Semihard processes can be explained by the QCD models such  as
\cite{Lev,Nas} etc.  The main question considered in this work is
whether one can apply the QGSM to the analysis of semihard processes
if one include the colour interaction between  colliding quarks
(antiquarks) and diquarks (quarks)
before the creation and dacay of $q\bar q$
(or $q(qq)$) string.

Our approach to the analysis of soft and semihard  hadron processes
in the QGSM takes into account  the dependence  of quark
distributions in hadrons and the quark fragmentation functions on
the transverse momentum $p_t$.  In the framework of the QGSM the
colour interaction of valence quarks and diquarks and sea quarks
(antiquarks) of colliding hadrons has been taken into account. This
one has been  calculated as the exchange by one nonpertubative gluon,
i.e., the cut-off parameter in the gluonic propagator has been
included.  The method  proposed to include the $p_t$-dependence in
the QGSM is close to the successive division of the transverse
momentum $p_t$ and $x$ between $n$-Pomeron showers  or $2n$ quark-antiquark
chains considered in our early works \cite{11,Lyk}. This method shows a
strong dependence of hadron characteristics on $n$.

The QGSM modified by such a way has allowed us to analyze the
correlation $<p_t>(N)$ and inclusive hadron spectra produced in
hadron collisions at transverse momenta up to $3-4$ Gev/c (Figs.
1-5). The calculations reproduce the data on  $<p_t>(N)$ at small
multiplicities $N$.  To describe the data at large $N$ one need, as
was mentioned above, to take into account the graphs of enhanced type
of Reggeon theory [2]. Inclusion of such diagrams result in the
appearance of $1/x$ terms in the distributions for sea quarks, which
give especially large contribution for $x\simeq 0$. Modified QGSM
can reproduce a lot of data about
invariant cross sections versus $p_t$, for example
of $\pi ^{\pm}$ mesons produced in $pp$ collisions at different values of x.

In the framework of new version of QGSM, the description of the
differential cross sections of $D^+$, $D^-$, $D^0$ and $\bar {D}^0$
mesons versus transverse momentum $p_t$ has been performed.
The comparison of the calculations with data show good agreement for
the case when Regge $\Psi $-trajectory has intercept, $\alpha _{\Psi
}(O)=0$. Such intercept value corresponds to the nonlinear $\Psi
$ trajectory. Main uncertainties of the calculations comes from poor
knowledge of intercept of $\Psi $-trajectory. At high  energies, the
predictions for $\alpha _{\Psi}(0)$ differ by a factor close to $3$.
Unfortunately , the large error in charm cross section measurement at
$\sqrt{s}=630$ GeV does not allow one to extract a useful constant on
$\alpha _{\Psi}(0)$.

The authors would like to thank Yu.A. Budagov for support and
constant interest to this work, A.B.Kaidalov, O.I Piskunova and
K.A.Ter-Martirosyan for useful discussions.

\section{Appendix}

Show if the function $D^2(q_{t}^2)$ can be approximated in form:
\begin{equation}
D^2(q_t^2)= \sum_{i=1}^{N}a_iexp(-b_iq_t^2),
\end{equation}
where  $a_i, b_i$-some parameters, then the integration of (10) is performed
by analitically. Consider firstly the simple case $n=1$, then  substituting
(21) and (6) to (2) we have:
\begin{equation}
g_{\tau}^{(1)}(\vec{k}_{1t}) = {\gamma \over \pi }\sum_{i=1}^Na_i
\int exp(-\gamma k_{1t})exp(-b_iq_{t}^2) \\
\delta^{(2)}(\vec{k}_{t}-\vec{k}_{1t}-\vec{q}_{t})d^{2}k_{1t}
d^{2}q_{t}.
\end{equation}
Integrating (22) over $d^{2}q_{t}$ and $d^{2}k_{1t}$ we get:
\begin{equation}
g_{\tau}^{(1)}(k_{t}^2)= \sum_{i=1}^{N}{a_i \over b_i}d_iexp(-d_ik_{t}^2),
\end{equation}
where $d_i={b_i\gamma \over {b_i+\gamma}}$
Substituting now
(1) and (16) to (10), we get the following expression for
$F^{(1)}_{\tau}(x_{\pm },\vec{p}_{t};x_{1,2}) $ :
\begin{equation}
F^{(1)}_{\tau}(x_{\pm },\vec{p}_{t};x_{1,2}) =
{f}^{(1)}_{\tau }(x_{1,2})G_{\tau \rightarrow h}(z,\vec{p}_{t})  \\
\int g_{\tau}^{(1)}(k_{t}) \tilde g_{\tau \rightarrow h}(\vec{p}_{t}-
z\vec{k}_{t})d^{2}k_{t}
\end{equation}
Substituting now
(23) and (17) to (24) we integrate it over
$d^{2}k_{t}$ and finally  have:
\begin{equation}
F^{(1)}_{\tau}(x_{\pm },\vec{p}_{t};x_{1,2}) =
{f}^{(1)}_{\tau }(x_{1,2})G_{\tau \rightarrow h}(z,\vec{p}_{t})
I_{1}(z,\vec{p}_{t}),
\end{equation}
where:
\begin{equation}
I_{1}(z,\vec{p}_{t})= \sum_{i=1}^{N}{a_i \over b_i}\gamma_{zi}
exp(-\gamma_{zi}p_{t}^2),
\end{equation}
here
 $\gamma_{zi}={d_i\tilde \gamma \over {d_i+z^2\tilde \gamma}}$
But for any n (10) is integrated over $d^{2}k_{t}$ anologous to this simple
case.

The question arises whether $D^2(q_{t}^2)$ (see (3)) can be represented
in the form (18). This approximation can be true at the following parameters:
$N=5, a_1=0.31, a_2=0.333, a_3=0.172, a_4=0.155, a_5=0.026$;
and $b_1=9.294, b_2=2.78, b_3=3.619, b_4=0.882, b_5=184.$

\newpage
\centerline{FIGURE CAPTIONS}

Fig. 1. Dependence of $<p_t>$ on the number of charged particles,
$N_{ch}$. The curve shows our calculation for $\sqrt{s}=63$ GeV.
Symbols denote the experimental data for $\sqrt{s}=63$ GeV, rapidity
interval, $\vert y\vert = 2$ \cite{21}.

Fig. 2. Invariant cross section of $\pi ^+$ mesons produced in
reaction $pp\ra \pi ^+X$ at $p_{lab}=200$ GeV/c as a function
of $p_t$ for the values $x=0$, $0.3$ and $0.6$. Experimental data
are from Refs. \cite{22}.

Fig. 3. Invariant cross section of $\pi ^-$ mesons produced in
reaction $pp\ra \pi ^-X$ at $p_{lab}=200$ GeV/c as a function
of $p_t$ for the values $x=0$, $0.3$ and $0.6$. Experimental data
are from Refs. \cite{22}.

Fig. 4. The inclusive cross section of the  reaction
$pp\ra \pi ^0X$ versus transverse momentum at $\sqrt{s}=52.7$ GeV.
The data are from Ref.\cite{23}.

Fig. 5. Differential cross sections of the reactions
$pp\ra D^+X$ (a), $pp\ra D^-X$ (b), $pp\ra D^0X$ (c), $pp\ra \bar
{D}^0X$ (d) at the energy $\sqrt{s}=27.4$ GeV as a function of the
square $p_t^2$ of the transverse momentum. Solid curves correspond to
calculations with the intercept $\alpha _{\psi }(0)=0$. The data are
from Ref. \cite{24}.

\begin{thebibliography}{99}
\setlength{\itemsep}{-0.36\baselineskip}
\bi{1}
A.B. Kaidalov
{\sl Phys. Lett.} {\bf B116} (1982) 459.
\bi{2}
A.B. Kaidalov, K.A. Ter-Martirosyan
{\sl Phys. Lett.} {\bf B117} (1982) 247; {\sl Yad.Fiz.} {\bf
39} (1984) 1545 and {\bf 40} (1984) 211.
\bi{3} Yu.M. Shabelsky
{\sl Yad. Fiz.} {\bf 44} (1986) 186.
\bi{4}
A. Capella, J. Tran Than Van
{\sl Phys. Lett.} {\bf B114} (1982) 450; {\sl Z.Phys.} {\bf C10}
(1981) 249
\bi{5} P. Aurenche, F.W.Boop, J.Ranft {\sl Phys. Lett.}
{\bf B114} (1982) 363; {\sl Z.Phys.} {\bf C13} (1982) 205.
{\sl Z.Phys.} {\bf C13} (1982) 205
\bi{6} X. Artru {\sl Phys. Rep.} {\bf 97} (1983) 33
\bi{7} G. t'Hooft {\sl
Nucl.Phys.} {\bf B72} (1974) 461
\bi{8} G. Veneziano {\sl Phys.Lett.}
{\bf B52} (1974) 220
\bi{9} M. Ciafoloni, G. Marchesini, G.Veneziano
{\sl Nucl.Phys.} {\bf B98} (1975) 472
\bi{10}
A.I. Veselov, O.I. Piskunova, K.A. Ter-Martirosyan
{\sl Prepr.ITEP} {\bf 176} (1984) Moscow
\bi{11}
G.I. Lykasov, M.N. Sergeenko
{\sl Z. Phys.} {\bf C52} (1991) 635; \\
{\sl Sov. J. Nucl. Phys.} {\bf 54(6)} (1991) 1037
\bibitem{Lyk}
G.I. Lykasov, M.N. Sergeenko
{\sl Z. Phys.} {\bf C56}
(1992) 697; \\
{\sl Sov. J. Nucl. Phys.} {\bf 55(9)} (1992) 1393
\bibitem{Ranft} J.Ranft {\sl Z.Phys.} {\bf C27} (1985) 413  \\
D. Perterman, J. Ranft, F.W.  Boop, {\sl Z. Phys.} {\bf C54} (1992)
682
\bibitem{Lev}
L.V.Gribov, E.M. Levin, M.G.Ryskin
{\sl Phys. Rep.} {\bf 100} (1983) 1
\bibitem{Nas}
P.Nasson, S.Dawson, R.K.Ellis
{\sl Nuc. Phys.} {\bf B303} (1988) 607
\bi{12}
F.E. Low
{\sl Phys. Rev.} {\bf D12} (1975) 163
\bi{13}
S. Nussinov
{\sl Phys. Rev. Lett.} {\bf 34} (1975) 1268
\bi{14}
D.G. Richards
{\sl Nucl. Phys.} {\bf B258} (1985) 267
\bi{15}
P.V. Landschoff and O. Nachtmann
{\sl Z. Phys.} {\bf C35} (1987) 405
\bi{16}
A. Donnachie and P.V. Landschoff
{\sl Nucl. Phys.} {\bf B311} (1988/89) 509
\bi{17}
J.R. Cudel and D.A. Ross
{\sl Nucl. Phys.} {\bf B359} (1991) 247
\bi{18}
F. Halzen, G.I. Krein and A.A. Natale
{\sl Preprint Univ. of Wisconsin-Madison} {\bf MAD/PH/702} (1992);
{\sl Phys. Rev.} {\bf D47} (1993) 295
\bi{19}
J.M. Cornwall {\sl Phys. Rev.} {\bf D26} (1982) 1453
\bi{20}
K.A.Ter-Martirosyan
{\sl Phys. Lett.} {\bf B44} (1973) 377
\bi{21}
W. Kittel, in:: Proc. of the XXIV Int. Conf. HEP,  edited by
R.Kotthaus and J.H.Kuhn, Munich, 1988, P.625.
\bi{22}
A.E. Brenner et al. {\sl Phys. Rev.} {\bf D26} (1982) 1479; \\
P. Capiluppi et al. {\sl Nucl Phys.} {\bf B70} (1974) 1
\bi{23}
F.W. Busser et al. {\sl Phys. Lett.} {\bf B46} (1973) 471
\bi{24}
M. Aguilar-Benitez et al. {\sl Phys. Lett.} {\bf B201} (1988) 176;  \\
{\sl Z. Phys.} {\bf C40 } (1988) 321.
\end{thebibliography}
\end{document}